\documentclass[conference]{IEEEtran}
\IEEEoverridecommandlockouts
\usepackage[utf8]{inputenc} % allow UTF-8 input for most chars
\usepackage{cite}
\usepackage{amsmath,amssymb,amsfonts}
\usepackage{algorithmic}
\usepackage{graphicx}
\usepackage{textcomp}
\usepackage{xcolor}
\usepackage{hyperref}
\hypersetup{
    colorlinks=true,
    linkcolor=blue,
    citecolor=blue,
    urlcolor=blue
}

\usepackage{listings}
\def\BibTeX{{\rm B\kern-.05em{\sc i\kern-.025em b}\kern-.08em
    T\kern-.1667em\lower.7ex\hbox{E}\kern-.125emX}}

\DeclareUnicodeCharacter{266F}{\#} % ♯ -> \#
\DeclareUnicodeCharacter{00B5}{\ensuremath{\mu}} % µ -> \mu
\DeclareUnicodeCharacter{2013}{--} % en-dash –
\DeclareUnicodeCharacter{2212}{-}  % minus sign − (different from ASCII -)

\begin{document}

\title{A Thermal Modeling Toolkit for Continuous-Wave Gaussian Second-Harmonic Generation in KTP Crystal}

\author{\IEEEauthorblockN{1\textsuperscript{st} Mostafa M. Rezaee}
  \IEEEauthorblockA{\textit{Department of Data Science} \\
    \textit{Bowling Green State University}\\
    Bowling Green, OH, USA \\
    mostam@bgsu.edu}
  \and
  \IEEEauthorblockN{2\textsuperscript{nd} Mohammad Sabaeian}
  \IEEEauthorblockA{\textit{Department of Physics} \\
    \textit{Shahid Chamran University of Ahvaz}\\
    Ahvaz, Khuzestan, Iran \\
    sabaeian@scu.ac.ir}
  \and
  \IEEEauthorblockN{3\textsuperscript{rd} Alireza Motazedian}
  \IEEEauthorblockA{\textit{Department of Physics} \\
    \textit{University of New Hampshire}\\
    Durham, NH, USA \\
    alireza.motazedian@unh.edu}
  \and
  \IEEEauthorblockN{4\textsuperscript{th} Fatemeh Sedaghat Jalil-Abadi}
  \IEEEauthorblockA{\textit{Department of Energy Engineering and Physics} \\
    \textit{Amirkabir University of Technology}\\
    Tehran, Iran \\
    f-sedaghat@aut.ac.ir}
  \and
  \IEEEauthorblockN{5\textsuperscript{th} Mohammad Ghadri}
  \IEEEauthorblockA{\textit{MIAE Department} \\
    \textit{Concordia University}\\
    Montreal, QC, Canada \\
    mohammad.ghadri@mail.concordia.ca}
}

\maketitle

\begin{abstract}
  We release an open-source finite-difference toolkit for computing temperature fields in continuous-wave (CW) second-harmonic generation (SHG) using potassium titanyl phosphate (KTP) crystals under Gaussian end-pumping. The toolkit includes modules for geometry and material definitions, boundary and cooling models, and transient and steady-state finite-difference solvers. Users provide beam and crystal parameters along with cooling profiles, and the solver returns spatiotemporal temperature fields with radial and axial profiles as exportable datasets. This implementation consolidates previous work into a single versioned repository with reproducible pipelines and parameterized scenario sweeps covering temperature-dependent versus constant conductivity, convection with or without radiation, and heat-transfer coefficients from $6.5$ to $2.0 \times 10^{4}$~W~m$^{-2}$~K$^{-1}$. The compiled Fortran kernels include built-in benchmark reporting. Validation is performed by reproducing published temperature distributions and trends for KTP under Gaussian CW pumping. The code is available as an open-source GitHub repository and is released under the MIT license as version v1.0.0, with an archived release on Zenodo identified by DOI 10.5281/zenodo.17266421 for citation and long-term access.
\end{abstract}

\begin{IEEEkeywords}
  Second-harmonic generation (SHG), continuous-wave (CW), potassium titanyl phosphate (KTP), heat equation, finite difference method (FDM), thermal lensing, open-source toolkit.
\end{IEEEkeywords}

\section{Introduction}

\IEEEPARstart{G}{reen} laser sources play a central role in modern photonics \cite{zu2024optical}, enabling applications across biomedical imaging \cite{kumar2014thermal, jung2022optical}, quantum communication \cite{bell2020digital, liu2024efficient}, material processing \cite{hu2013engineered, liu2023high}, remote sensing \cite{rodier2011calipso, szafarczyk2022use}, and precision metrology \cite{nomura2019precision}. Direct green-emitting gain media remain limited in power and beam quality \cite{feng2023simple}, so frequency doubling of near-infrared lasers---commonly at 1064~nm from Nd-based systems---has become the dominant approach for generating high-power, narrow-linewidth green output. Second-harmonic generation (SHG) is typically implemented in nonlinear crystals such as potassium titanyl phosphate (KTP), which offers favorable thermal, optical, and mechanical properties. However, achieving stable and efficient SHG at high continuous-wave (CW) power levels \cite{ricciardi2010cavity} requires careful thermal management \cite{xu2005110} because absorption-induced heating within the crystal degrades phase-matching conditions and distorts the optical mode \cite{sahm2011thermal, zheng2001influence}.

In practice, heat deposited by absorption in the nonlinear crystal drives temperature rise, thermal lensing, and slow drift of phase mismatch \cite{xu2005influence}. For KTP \cite{zhou2023hydrothermal} under Gaussian end-pumping, two modeling choices often dominate accuracy: treating thermal conductivity as temperature-dependent rather than constant, and using realistic boundary and cooling conditions (convection with possible radiative exchange) instead of oversimplified fixed-temperature faces. Prior studies show that temperature-dependent conductivity can shift peak temperatures by tens of kelvin \cite{van2023effect}, and radiation becomes relevant primarily for large radii and beam sizes. Meanwhile, the wide range of reported convection coefficients complicates comparisons across labs, making reproducible benchmarking difficult \cite{sabaeian2015temperature}.

Here, we describe a consolidated open-source finite-difference toolkit for KTP thermal behavior in CW Gaussian SHG with standardized inputs, solvers, and outputs. The repository packages an Intel Fortran finite-difference implementation with ready-to-run examples, reference result files, citation metadata, and licensing. A versioned release archived on Zenodo enables reproducibility and long-term access \cite{sabaeian_17266421}.

Users provide beam and crystal geometry along with cooling parameters, and the solver returns spatiotemporal temperature fields with axial and radial profiles suitable for downstream SHG analysis. Scenario sweeps cover temperature-dependent versus constant conductivity, convection alone versus convection plus radiation, and convection coefficients spanning laboratory air to high-conductance mounts---mirroring ranges reported in the literature for KTP. Reference runs reproduce the continuous-wave Gaussian KTP study, including sensitivity to conductivity models and boundary conditions, allowing users to verify their environment before exploring new setups \cite{sabaeian2015temperature}.

The toolkit is designed for users who need transparent baselines and quick checks before exploring alternate crystals or boundary models. This includes researchers comparing thermal models and reporting reproducible benchmarks, educators who need transparent and modifiable code for classroom labs, and engineers sizing crystals and mounts with explicit cooling assumptions.

\section{Toolkit Lineage}

This toolkit unifies prior heat-equation studies into a single, versioned codebase, reproducing the continuous-wave (CW) Gaussian KTP results \cite{sabaeian2015tempdist} and employing the analytical heat-equation solution as its foundational baseline \cite{sabaeian2008analytical}. This implementation became the stable anchor layer for our later works. Building on this foundation, separate, task-specific toolkits and articles were developed that consume the exported temperature fields—covering pulsed heating, thermally induced phase mismatch, CW double-pass efficiency under heating, non-Gaussian source models, and heat-coupled CW formulations.

An early study \cite{sabaeian2015tempdist} extended the analytical framework to include temperature-dependent thermal conductivity and radiative boundary conditions in a Gaussian end-pumped KTP crystal, capturing more realistic thermal transport behavior. This was followed by a comprehensive numerical approach for pulsed excitation \cite{rezaee2015complete}, which solved the complete anisotropic, time-dependent heat equation under repetitively pulsed Gaussian beams. Building upon that, the spatiotemporal thermal phase mismatch effects were formulated and analyzed in detail \cite{rezaee2015thermally}, demonstrating how accumulative heating and temporal pulse spacing influence refractive-index dynamics.

Subsequent works focused on CW and nonlinear coupling effects. In \cite{sabaeian2015temperature}, a depleted-wave model was proposed to study temperature-induced efficiency losses in double-pass type-II SHG cavities under Gaussian beams. Non-Gaussian source behavior was later introduced through pulsed Bessel–Gauss beam modeling \cite{sabaeian2014pulsed}, extending the toolkit to broader beam families and validating the nondepleted approximation limits for high-intensity short pulses. Later, a fully coupled formulation integrating the heat, SHG, and thermally induced phase-mismatch equations was established in \cite{sabaeian2014heat}, simultaneously treating thermal lensing and TIPM effects in a CW double-pass configuration.

Together, these studies represent a cohesive progression from the analytical baseline toward a modular, physics-grounded computational toolkit suite capable of addressing a wide range of laser–crystal thermal interactions—from steady-state Gaussian heating to complex, nonlinear, and time-dependent regimes.

\section{Toolkit Architecture}

\subsection{Scope}
The toolkit is a Fortran finite-difference implementation of the transient and steady heat equation in a KTP crystal under Gaussian continuous-wave pumping with temperature-dependent thermal conductivity and Robin (convection with or without radiation) boundary conditions. It targets reproducible temperature fields $T(r, z, t)$ and steady-state profiles for use in downstream SHG analyses.

\subsection{Modules}
The Core Solver solves the long-transient heat equation in cylindrical coordinates on an $r$--$z$ grid with a Gaussian source proportional to
\begin{equation}
  \exp\left(-\frac{2r^2}{\omega_0^2}\right)\exp(-\alpha z),
\end{equation}
and temperature-dependent conductivity $K(T)$. Steady state emerges as $t \to \infty$. The solver is implemented in Intel Fortran (IFORT) on Linux. Geometry and Materials defines a cylindrical KTP of radius $a$ and length $l$ with material set $\{\rho, c, K_0, \alpha, \epsilon\}$, where conductivity follows $K(T)$ as specified in \cite{sabaeian2015tempdist}. Boundary and Cooling implements convection with heat-transfer coefficient $h$ spanning reported values---$6.5$ to $50$~W~m$^{-2}$~K$^{-1}$ for air and approximately $15{,}000$ to $20{,}000$~W~m$^{-2}$~K$^{-1}$ for high-conductance mounts or water---optionally plus radiation via
\begin{equation}
  \epsilon\sigma(T^4 - T_s^4).
\end{equation}
Radiation is negligible for small radii and beam sizes but becomes significant for larger surfaces and spots. Studies and Reporting provides packaged examples that produce published trends for KTP under CW Gaussian pumping, comparing constant versus temperature-dependent $K$ with and without radiation, as well as $h$ sweeps. Output datasets include axial, radial, and transverse temperature profiles.

\subsection{Data Flow}
Inputs include crystal geometry $(a, l)$, pump power and spot size $(P, \omega)$ with absorption coefficient $\alpha$, material properties $(\rho, c, K(T))$, and boundary parameters $(h, \epsilon, T_\infty, T_s)$ taken from the reference configurations. Intermediates consist of time-indexed temperature fields $T(r, z, t)$ and steady-state one-dimensional profiles along $r$ and $z$ for validation. Outputs are plain-text profiles in \lstinline{results/}, such as \lstinline{ST_085_time_1_T_r.plt}, \lstinline{..._T_t.plt}, and \lstinline{..._T_z.plt}.

\subsection{Interfaces}
Users interact with the toolkit through source code that compiles to a Fortran binary executable. Build instructions are provided in Section~V.

\section{Validation and Benchmarks}

We solved the CW–Gaussian KTP temperature-distribution problem of \cite{sabaeian2015tempdist} with this toolkit. Figures and parameters mentioned below are taken from the reference paper \cite{sabaeian2015tempdist}.

\subsection{Representative Replications}
\begin{enumerate}
  \item \textbf{Temporal center point (Fig.~4).} We validated the solver using a KTP cylinder with $a = 2$~mm and $l = 20$~mm under a Gaussian pump of $P = 80$~W and $\omega = 100$~$\mu$m, with convection and radiation boundaries, comparing constant $K$ versus temperature-dependent $K(T)$. Building and running the Fortran solver (see Section~IV) extracts the on-axis entrance temperature as a function of time, reaching steady state as $t \to \infty$. The toolkit reproduces the reported $\sim 69$--$70$~K higher center temperature for $K(T)$ versus constant $K$ at 80~W. The full transient to steady state takes approximately 360~s on an Intel i5 2.53~GHz/4~GB Linux system using IFORT.

  \item \textbf{Radial and axial steady-state profiles (Figs.~5--6).} Using the same geometry and pump parameters, we evaluated $T(r, z = 0)$ and $T(r = 0, z)$. After convergence, the radial and axial slices match the shape and extrema of the published curves, with the lateral boundary correctly returning to 300~K at $r = a$. The results confirm higher temperatures under $K(T)$ with proper recovery to 300~K at the cooled lateral surface, and axial profiles match the higher baseline under $K(T)$. Runtime is comparable to Fig.~4.

  \item \textbf{Cooling-model sensitivity (Figs.~7--13).} We swept convection coefficients $h$ over reported values (air: $6.5$--$50$~W~m$^{-2}$~K$^{-1}$; high-conductance/water: $15{,}000$--$20{,}000$~W~m$^{-2}$~K$^{-1}$) and optionally toggled radiation. Running steady-state cases and comparing entrance-face temperatures shows insensitivity across $h = 6.5$--$50$~W~m$^{-2}$~K$^{-1}$ and approximately 20~K lower temperature for $h = 15{,}000$--$20{,}000$~W~m$^{-2}$~K$^{-1}$ versus air. Radiation is negligible for $a = 2$~mm and $\omega = 100$~$\mu$m even at higher powers, but for $a = 5$~mm, $\omega = 3$~mm, and $P = 530$~W, radiation reduces temperature by approximately 15~K at the input facet and 10~K at the output facet.
\end{enumerate}

\subsection{Sanity Check}
Read \lstinline{ST_085_time_1_T_r.plt} and confirm the final radial sample at $r = a$ equals 300~K (ambient), consistent with the fixed lateral boundary condition $T(r = a, \cdot, \cdot) = 300$~K. This is a quick file check and requires no recompilation.

\section{Usage and Reproducibility}

\subsection{Environment}
The toolkit requires Intel Fortran (IFORT) on Linux or WSL. The main source code is located in \lstinline{src/Code_SHG-CW-G-Heat-Equation.f90}, and reference outputs are provided in \lstinline{results/}. The solver is deterministic for a grid of $200 \times 150$ nodes with time step $\Delta t = 6.97 \times 10^{-6}$~s and uses no random seeds.

\noindent To build and execute the solver in a single command, use:

\begin{lstlisting}[language=bash, basicstyle=\small\ttfamily, frame=single, breaklines=true, breakatwhitespace=true]
ifort src/Code_SHG-CW-G-Heat-Equation.f90 -o shg_cw_heat && ./shg_cw_heat
\end{lstlisting}

\noindent After the solver completes, compare the produced \lstinline{.plt} files with the packaged reference files in \lstinline{results/}, such as \lstinline{ST_085_time_1_T_r.plt}, \lstinline{..._T_t.plt}, and \lstinline{..._T_z.plt}.

\section{Limitations and Roadmap}

\subsection{Current Scope and Limitations}
The solver addresses the scalar heat equation in a cylindrical, azimuthally symmetric KTP crystal under a CW Gaussian source. It does not compute optical SHG fields or vectorial effects---the $\phi$-dependence is dropped by symmetry. The implementation supports CW Gaussian heating only, with transients integrated until steady state. Pulsed excitation and non-Gaussian sources are not included in this release. Property tables and validation target KTP; other nonlinear crystals mentioned in the background are not parameterized here. The end faces use convection with or without radiation boundary conditions, while the lateral surface is fixed at 300~K. More general lateral boundary models, non-axisymmetric configurations, and full three-dimensional geometries are beyond the current scope.

\subsection{Typical Failure Modes}
Several common modeling errors can lead to significant inaccuracies. Treating $K$ as constant when temperature-dependent $K(T)$ is required leads to large underestimation of peak temperatures---reported differences reach approximately 70~K at 80~W. Assuming radiation is negligible in large-radius, large-spot, high-power cases mispredicts entrance and exit-face temperatures by tens of kelvin. Using convection coefficients $h$ outside published ranges (air: $6.5$--$50$~W~m$^{-2}$~K$^{-1}$; high-conductance: approximately $15{,}000$--$20{,}000$~W~m$^{-2}$~K$^{-1}$), or mixing boundary models across comparisons, breaks reproducibility. Finally, choosing grid and time-step values that violate numerical stability produces non-convergent or biased results. This implementation uses $200 \times 150$ nodes with $\Delta t = 6.97 \times 10^{-6}$~s.

\subsection{Near-Term Roadmap}
Future development plans include extending the heat source to pulsed operation and regenerating the transient validations using the pulsed studies cited herein. Additional material parameter sets will be added for other nonlinear crystals (LBO, BBO, Mg:PPLN) with matched examples. Documentation for post-processing hooks will be provided to couple exported $T(r, z, t)$ profiles to phase-mismatch and thermal-lensing analyses.

\section{Availability and License}

\subsection{Availability}
Source code: \href{https://github.com/Second-Harmonic-Generation/SHG-CW-G-Heat-Equation}{GitHub repository} (release tag \texttt{v1.0.0}); archived at Zenodo (DOI: 10.5281/zenodo.17266421).

\subsection{License}
This project is released under the MIT License, as detailed in the \lstinline{LICENSE} file in the repository.

\section{Conclusion}
We have presented an open-source, versioned toolkit that consolidates our prior heat-equation work into a single Intel Fortran implementation with standardized inputs and outputs, packaged reference runs, and archival release. The approach was validated by reproducing published temperature distributions for CW Gaussian pumping in KTP, shipping axial and radial profile files that enable direct checks of conductivity models and boundary and cooling assumptions. This provides a reproducible starting point for benchmarking environments, comparing constant-$K$ versus $K(T)$ cases, and performing documented $h$-sweeps. Concrete extensions were outlined, including pulsed sources, broader crystal parameter sets, and documented post-processing hooks. Together, these results establish a coherent foundation for future studies of thermally coupled SHG and related heat-conduction problems.

\bibliographystyle{IEEEtran}
\bibliography{references}

\end{document}